# The Promise and Prejudice of Big Data in Intelligence Community

Karan Jani[1]

*Sam Nunn Security Program*
Sam Nunn School of International Affairs
Georgia Institute of Technology
October 26, 2016

## ABSTRACT

Big data holds critical importance in the current generation of information technology, with applications ranging from financial, industrial, academic to defense sectors. With the exponential rise of open source data from social media and increasing government monitoring, big data is now also linked with national security, and subsequently to the intelligence community. In this study I review the scope of big data sciences in the functioning of intelligence community. The major part of my study focuses on the inherent limitations of big data, which affects the intelligence agencies from gathering of information to anticipating surprises. The limiting factors range from technical to ethical issues connected with big data. My study concludes the need of experts with domain knowledge from intelligence community to efficiently guide big data analysis for timely filling the knowledge gaps. As a case study on limitations of using big data, I narrate some of the ongoing work in nuclear intelligence using simple analytics and argue on why big data analysis in that case would lead to unnecessary complications. For further investigation, I highlight cases of crowdsource forecasting tournaments and predicting unrest from social media.

---

[1] kpj@gatech.edu

# Table of Contents

# 1. Introduction

*"When a person monitors a database, it's small data, & when a database monitors a person, it's called big data."*

**-- A Quora User**

We live in a knowledge economy, where data is the new currency. The amount of data interpreted per day through an average smartphone user exceeds the text data of all combined 20th century English literature. Data has been central to the physical sciences, which progressed through painstaking analysis of controlled experiments. Most of the standard data analysis had been invented in the process of solving a complicated numerical problem (example: Monte Carlo). The analysis of large, complex data sets in academia has led to discovery of Higgs Boson to decoding of geome. In industry, the dependence on data increased only post dot com revolution. Currently, the asset of all IT giants is their user database (Google, Facebook). The analysis tools adapted by industries are the ones first developed in academia.

For a nation's Intelligence Community (IC), data and its analytics have been central for its mission to anticipate surprise. However, up to recently the conventional approach of IC was to most rely on human judgement to decode databases and find trends. The boom in open source database from industries since last decade, has led the IC to aggressively revamp its infrastructure and R&D and be at par with the data analytics standards of industry and academia (see figure-1). The motivation behind this study is to investigate the scope and limitations of adapting data sciences in intelligence gathering.

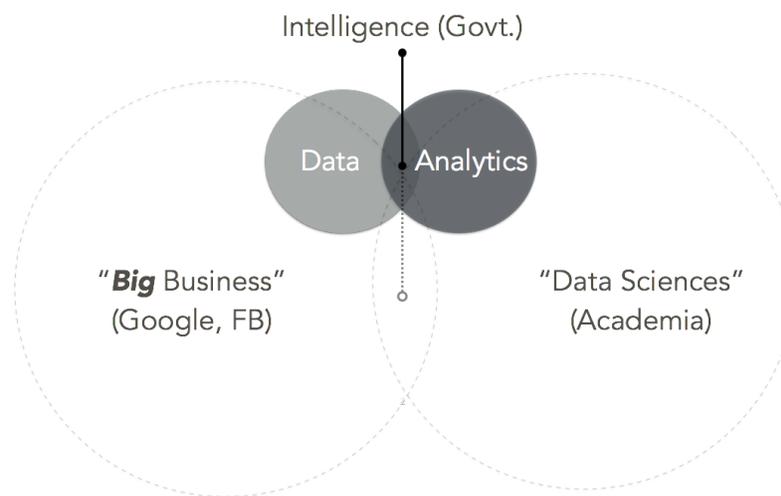

**Figure 1:** A venn diagram overlap of the areas in big data between academia, industry and government.

## 1.1 Defining the "Big Problem" in a Data

The first instance where the "big data problem" is defined in the literature is a 1997 paper by NASA Ames Research Center:

- *"Visualization provides an interesting challenge for computer systems:* **data sets are generally quite large, taxing the capacities of main memory, local disk, and even remote disk.** *We call this the problem of big data"*[2]

The issue defined by NASA researchers was less on the complexity of the dataset itself, but more on the visual interpretation of the data, which demands a better technological solution. The dataset they were dealing with was of fluid mechanics, which had a known solution under good approximation, but the visualization demanded interaction between each fluid element to form the clumps and the turbulence.

The understanding of "big data" has evolved significantly since then, but the structure of every definition still follows the one by NASA. The most standard definition of big data used in the literature is given McKinsey Global Institute[3]:

- *"Datasets whose size is beyond the ability of typical database software tools to* **capture, store, manage, and analyze**"

This definition is robust and invariant to any complexity of data set there may be. In some sense it captures the universal "big data problem".The four highlighted issues in this definition (*capture*: acquiring the database, *store:* memory of the database, *manage*: transferring the database, and *analyzing* the database), interrelate to create a practical challenge for solving big data problem. This is best captured in what is known in the literature as the "Vs" of big data (see Figure-2)[4]:

- **Volume**
  - This fundamentally defines the bigness of "big data". The total open source data in 2000 was $\sim 10^{20}$ bytes. By 2020 it is expected to reach $\sim 10^{22}$.

---

[2] NASA Ames Research Center. *Application-Controlled Demand Paging for Out-of-Core Visualization,* 1997
https://www.nas.nasa.gov/assets/pdf/techreports/1997/nas-97-010.pdf

[3] McKinsey Global Institute, *Big Data: The next frontier for innovation, competition, and productivity Report*, June, 2012.

[4] Roberto V. Zicari. *Big Data: Challenges and Opportunities*. 2013
http://www.odbms.org/wp-content/uploads/2013/07/Big-Data.Zicari.pdf

- - Every day Twitter and Facebook generates $\sim 10^{13}$ bytes of data. (see figure-3)

- **Variety**

  - This describes the complexity of data types: it can be in form of videos, photos, web data.
  - Over 80% of data is unstructured and simply cannot be expressed in some rows and columns.

- **Velocity**

  - This describes the rate of acquiring data before one can start analysis. One ideally has to merge the new available dataset with the older dataset (say stock market), which can at times not be an easy stitch.

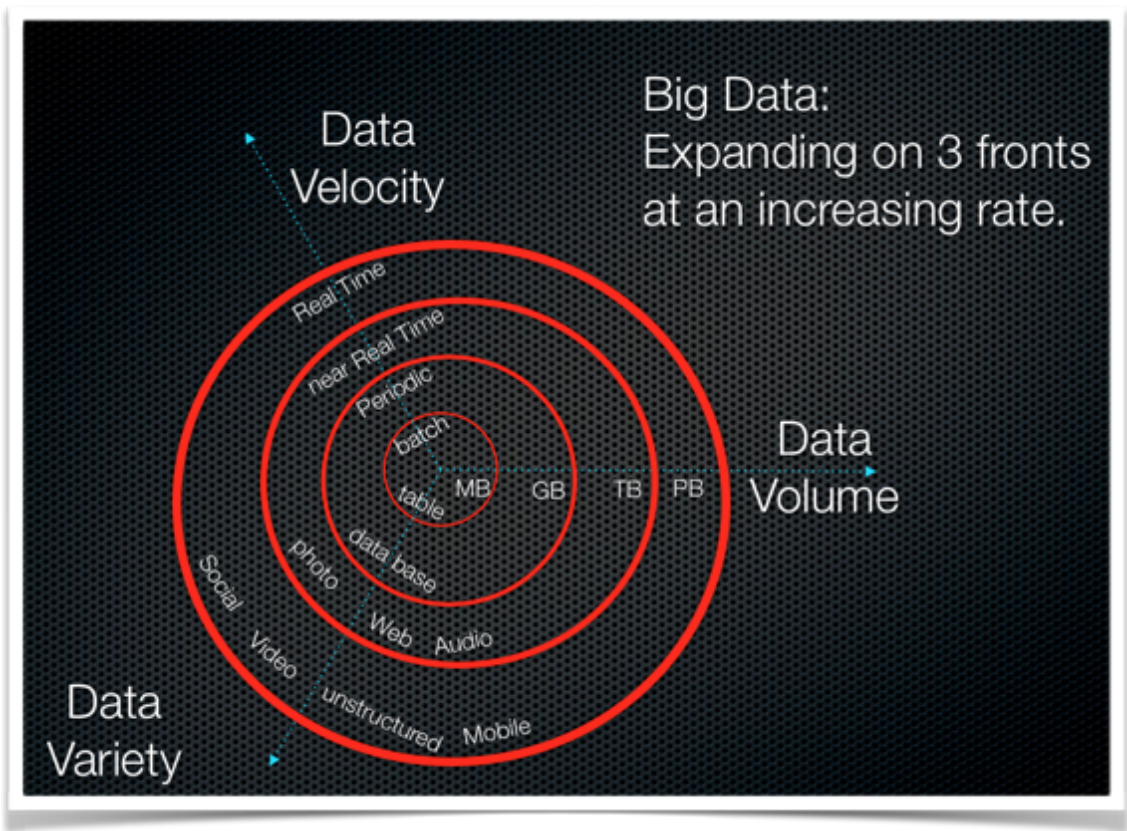

**Figure 2:** The conventional **V**s of big data. Each circle with increasing radius refers to increasing difficulty in analyzing the data.[5]

---

[5] Figure courtesy: Api.ning.com

http://api.ning.com/files/O6-JQcfS6sxRuZ8I2i5nJVa59xL-krT-a6UqeoLNaHwL2w-JSR-Cy56PmikOywRQgy2gDfYxLAb0Hs*VFr8IePv5QFBJdhDH/BigData.001.jpg

Beyond these conventional "**V**s" that define big data problem, in the case of intelligence community there are three more Vs that play a very crucial role[6]:

- **Veracity**

  - The accuracy and authenticity of dataset remains a big unknown. One cannot guarantee if what is available is a full dataset and if not it affects the prediction by IC.
  - The uncertainty in the database has been rising at the same rate as the size of the data itself. (See figure-3)

- **Volatility**

  - This refers to the latency by which one can analyze a given dataset. For time-sensitive information, the IC wants almost real time analysis.

- **Value**

  - The value of analyzing big data is usually the hardest predict early on. The value of a dataset may not be of any relevance for real time analysis but may be at times for looking at historical trends.

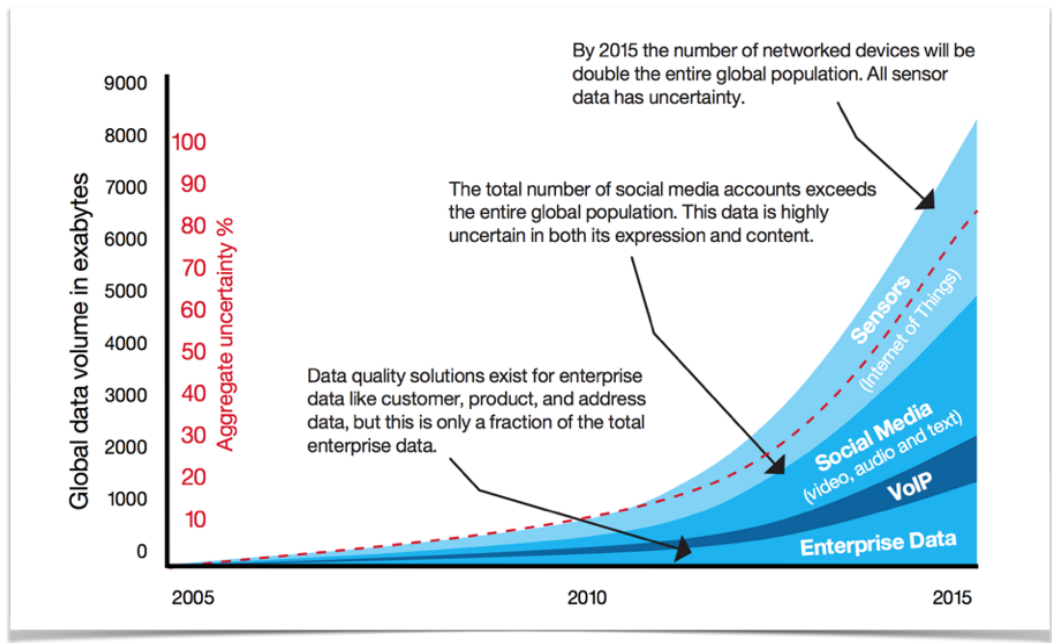

**Figure 3:** The exponential increase in global open source data in last decade. The dashed red line refers to the uncertainty (**V**eracity) of the data.[7]

---

# 2. Big Data and Intelligence Community

*"The IC's routine work of collection, processing, exploitation, and dissemination, and analysis is still largely organized on the Cold War model of seeking out sparse and secret information."*

**-- Paul B. Symon and Arzan Tarapore** [8]

The applications of big data sciences to Intelligence Community (IC) and subsequently National Security is evident. However, big data is one of those rare disruptive technologies that escalated in industry and academia at a much higher rate compared to defense research. The level of access to user data and sophisticated mining tools that are available with IT giants like Facebook, Google and IBM, far exceeds the scale at which current generation of infrastructure with the IC. At the same instance, the open source information that is available through social media has developed into to a new form of crucial intelligence: *SOCMINT* [9]. To timely utilize such massive open source data, the need for IC to catch up with the big data science has never been more urgent than now.

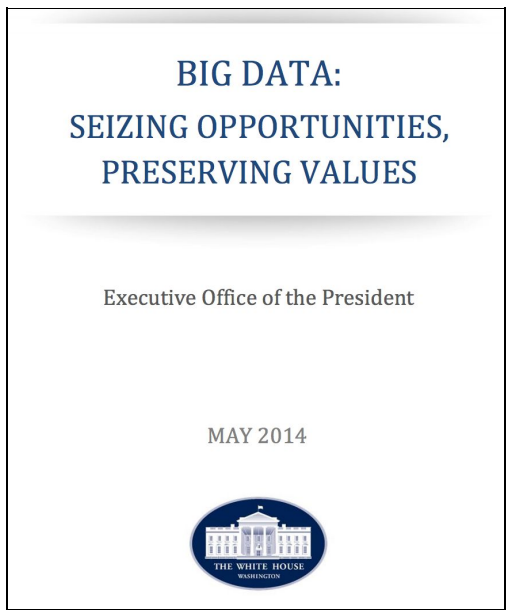

In recent years the United States government has placed highest importance in developing national strategy to address the big data problem. The White House released a report on Big Data[10] (left). Though it mainly focuses on the ethical issues and policy formation related to using big data across the spectrum, it lists some of the highlights of its usage in context of defense and intelligence community:

"During the most violent years of the war in Afghanistan, DARPA deployed teams of data scientists and visualizers to the battlefield… these teams .. fused satellite and surveillance data to visualize how traffic flowed through road networks, making it easier to locate and destroy improvised explosive devices"

---

[8] Paul B. Symon and Arzan Tarapore. *Defense Intelligence Analysis in the Age of Big Data.* Forum of Defense Intelligence and Big Data, 2015
http://ndupress.ndu.edu/Portals/68/Documents/jfq/jfq-79/jfq-79_4-11_Symon-Tarapore.pdf

[9] Center for Analysis of Social Media
http://www.demos.co.uk/project/intelligence/

[10] *Big Data: Seizing Opportunities, Preserving Values.* Executive Office of the President.
https://www.whitehouse.gov/sites/default/files/docs/big_data_privacy_report_may_1_2014.pdf

Based on recent statistics[11], the amount of data post 9/11 simply from the surveillance has increased by 1600%. There are over 7 million computing devices currently within the United States armed services, which will increase to 15 million by 2020. Some of the primary areas of national security where such big data is being generated and are related to the IC are[12]:

- Maritime Security
- Cyber Security
- Money Laundering
- Multi-INT Analysis
- Space Situational Awareness

The information analysis of IC, for any given form of data set is shown in Figure-4. It is a three stage structure and with each step the **V**alue of dataset increases (i.e. noise decreases). With the rise of big data, each conventional approach of IC to these three steps gets too time consuming.

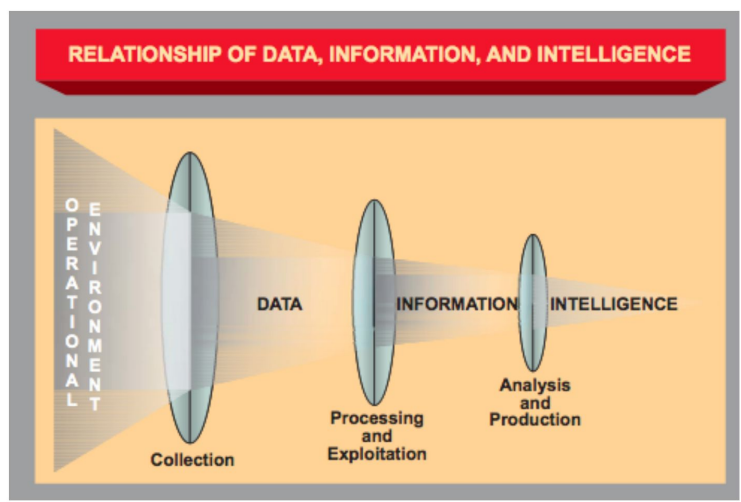

**Figure 4:** Schematic diagram describing the relation between big data and intelligence community. The three stages referring - collection, processing, analysis - increasingly cleans the data to make it more **V**aluable.[13]

[11] Chris Young. *Military Intelligence Redefined: Big Data in the Battlefield,* Forbes, March 2012
http://www.forbes.com/sites/techonomy/2012/03/12/military-intelligence-redefined-big-data-in-the-battlefield/#367ab43f718f

[12] Sean Fahey. Big Data Analytics and National Security, 2012
http://web.stanford.edu/group/mmds/slides2012/s-fahey.pdf

[13]Sean Fahey. Big Data Analytics and National Security, 2012
http://web.stanford.edu/group/mmds/slides2012/s-fahey.pdf

One hopes that by developing the tools that are already available in the standard big data analysis (industry+academia), the information processing in IC can benefit in three major ways:

(I) the collection and characterizing of data becomes automated.

(II) the processing time for analyzing complex data structure almost reduces to being real time.

(III) the presentation of result can be refined to only key features that led to effective decision making.

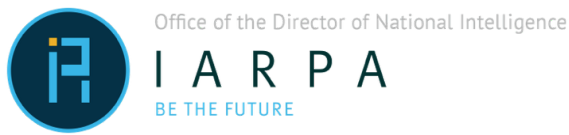 In the attempt to develop in-house expertise to tackle big data problem for the United States Intelligence Community, in 2006 the Intelligence Advanced Research Projects Activity (IARPA) was established.

Considered as "High risk, high payoff", and a startup type of environment, the four research areas, in line with big data, that are focused within IARPA are:

**i) Anticipating Surprise:** forecasting tournament

**ii) Incisive Analysis :** maximizing insights from dynamic datasets

**iii) Smart Collection:**

**iv) Safe & Secure Operations:** threat detection, quantum computing

One of the rares in the conventional intelligence community, IARPA heavily collaborates with industry, academia. IARPA also deals dominantly with open source data and engages general public in data sciences puzzles. Two of the most successful programs of IARPA in the open source sector has been,

**FOREST (Forecasting Science and Technology)**, which initiated crowdsourced, forecasting tournaments to predict disruptive technologies.

**OSI (Open Source Indicators),** where social media is analyzed to predict warnings about real world events.  This program has been credited for predicting Ebola much before the outbreak.

# 2. The Big Limitations of Big Data

*"If 50 years of research in artificial intelligence has taught us anything, it's that every problem is different, that there are no universally applicable solutions. An algorithm that is good at chess isn't going to be much help parsing sentences, and one that parses sentences isn't going to be much help playing chess* **solving problems will often require a fair amount of … "domain knowledge"—specific information about particular problems, often gathered painstakingly by experts.** *So-called machine learning can sometimes help, but nobody has ever, for example, built a world-class chess program by taking a generally smart machine, endowing it with enormous data, and letting it learn for itself…* **Big Data is a powerful tool for inferring correlations, not a magic wand for inferring causality**"

<div align="right">

--**Gary Marcus,** Professor of Cognitive Science, New York University[14]

</div>

By the very construct, data analytics algorithms (primarily machine learning) are designed to solve problems where we have a prior understanding of the complexities in data sets. Those complexities can be common across spectrums and multiple disciplines, hence one might be tempted to apply, for example, an algorithm used for face recognition on Flickr to identify black holes from data of Hubble Space Telescope. But the beauty of complex data lies in the degeneracy. Too many things will look too similar to form any coherent judgement. To break that degeneracy and find meaningful correlation in the data sets relies fundamentally on domain knowledge.

Just the like the "Vs" of big data, experts in the field of analytics have their own list and ranking of the limitations of big data sciences. But by and large, the five major issues of big data sciences, which are relevant for the discussion in this paper, can be clubbed as following[15]:

### a. Big data gives too many correlations

- If one looks at the trends in the criminal and financial data sets from 2006-2011, there exists a correlation between market share of Internet explorer and number of deaths in the United States. This of course is nonsensical but only because we have a prior knowledge that such correlation is of no use. However, in territories where we

---

[14] Gary Marcus. Steamrolled by Big Data. New Yorker, March 2013.

http://www.newyorker.com/tech/elements/steamrolled-by-big-data?utm_source=datafloq&utm_medium=ref&utm_campaign=datafloq

[15] Gary March & Ernest Davis. *Eight (No, Nine!) Problems With Big Data. New York Times,* April 2014
http://www.nytimes.com/2014/04/07/opinion/eight-no-nine-problems-with-big-data.html?_r=0

do not have the familiarity to make such judgements, we may well be easily carried away by the correlations from big data analysis.

## b. Big data can be gamed

- Every correlation that we see in big data is ultimately just a number. This number in most cases refers to peak of some probability distribution. But as soon as one understands the patterns in the data set, it is trivial to fake and hack the the data set such that every time you get the same correlation. Thus, big data analysis could be purposefully be misguided.

## c. Big data changes with time

- Most of the interesting data set (political support on social media, stock market, weather forecast), which rely big data analysis are dynamic and their trends fluctuates with time. The correlation obtained on the set using such analytics are valid only for a given time. Hence, all judgments from big data come with a finite deadline.

## d. Big data is less robust

- The hype on the success of big data mostly overlooks the fact that the problems it solves are rather common data sets. For more complex data sets, like whether can you predict translation of a word in other language, there has been no breakthroughs, though it does fall under same class as predicting behavior from social media response. Previously hailed as the big success of big data, the Google Flu Trends failed badly recently by " predicting more than double the proportion of doctor visits for influenza-like illness than the Centers for Disease Control and Prevention's surveillance" [16].

## e. Big data is only a tool

- Like the many technologies of past and present, big data is no more than just an analysis tool that gives a little further heads up than conventional approaches. At no instance we can write super codes that will predict every trend and come back with solutions to most complex issues. The dependence on big data from industries to defense is fair, but it cannot be measured with breakthroughs such as satellite transmissioning, nuclear reactors and vaccines.

---

# 5.1 Limitations to Intelligence Community

*"We simply cannot be the slaves to the data"*

**-- Anonymous**

### i) Big Data Needs Experts

- When faced with petabytes of unstructured data sets, scattered across multiple mediums and highly uncertain source of authenticity, one simply cannot use a brute force method and apply machine learning algorithms. To expedite the decision making process, experts in the field would have to downsample datasets and prioritize feasible targets. This requires familiarity with the context of the situation and the questions one wants to answer using data analysis.

- The primary need for correlations within complex, unstructured data sets is to guide IC in terms of planning and strategy. However, as noted in previous section, big data analysis can give multiple correlations, some of which are absurd if the domain knowledge is not available *a priori*. Blindly following the outputs from machine learning algorithms will ultimately lead to low-latency in decision making and wastage of key resources. The final step of the intelligence - big data cycle has to hence be the expert leadership

- Example: The Islamic State (IS) has been very active in using social media (Twitter) and YouTube to carry forward their agenda. Using such open source intelligence, along with the surveillance and intelligence from ground, in principle big data can correlate their possible location and resources . But it still does not provide any strategy to the IC for next course of action. In fact, IS has been wise in using twitter so that it is more populated with similar keywords, leading to unuseful correlations. It is at such places the IC needs human interference and experts.

### ii) Big Data Cannot Address Knowledge Gap

- The central usage of big data in IC is to bridge the information gaps in otherwise unstructured data sets. For example, say there are blurred surveillance pictures of a potential activity of national security concern, and, at first, completely unrelated posts on social media and internet from a particular territory. Machine learning algorithms may pick up trends between the such divergent data sets and provide a correlation which otherwise will go unnoticed or have high latency in the conventional approaches of IC. This will be a classic example of filling the information gap. However, what more does it say? If suppose this was the case for Syria, does this imply non-state actors are going to attack a western country in some finite time?  One cannot answer such questions at all without having experts in field on those region who can join the dots to predict potential actions by non-state actors. This inability of big data to predict causality is where lies the *knowledge gap*.

- Much of the knowledge gap is a acquired by understanding "long data" rather than big data[17]. *Long data* in this context refers to understanding a trends over a long period of time. The group project of this year Sam Nunn Security Program described how technology over last half century has transformed IC. Using this domain knowledge we can address if future disruptive technologies like augmented/virtual reality will lead to an effect similar to the one observed in Cold War era due to satellite communication. Big data by itself in no known analysis can meaningfully answer such questions.The information bridged by big data is only valid within a slice of time in knowledge cone.

- Example:
(i) During the Arab Spring, social media played a very important role in spreading on the movement across the region. However, with all the information at hand, none of the big data analysis available to IC at the time could predict the trajectory of the protest from Tunisia to Egypt, neither could have predicted the unrest following a 'successful' revolution of taking down authoritarian regimes. An expert instead could have predict the fate of these nations knowing about the politics and the history of the region.

(ii)The joint India-US remote sensing satellite, NISAR, will be monitoring the Indian Ocean with high resolution for predicting weather. Suppose these heavy database captures a new Chinese vessel with weapons that directly correlates with previously known information. This surely bridges the intelligence gap. But then what are the intentions of Beijing with such new weapons? The big data may not answer that but an expert in the field may point China's past strategies to intimate its neighbour for the claim of oil basins in Indian Ocean.

### iii) Big Data Analysis has Finite Lifetime

- All of current computational infrastructure and algorithms that enable us efficient implications of big data analytics has a lifetime which is proportional to the size of the data set. The data analysis tools we had available in mid 2000s have already been quashed with the exponential rise of open source data from social media. Hence it has to be acknowledged that the current investments IC is making in big data sciences will be dusted in less than 5 years from now with the rise of Internet of Things and cloud computing.

- For a long term plan, what the IC can at best do is develop a framework (along with interactive softwares) within which experts from varying intelligence fields can

---

efficiently conduct:
(a) processing, exploitation, and dissemination of complex data sets,
(b) the final stage of implementing strategies from the results of analysis.
This has to be thought through in a robust form just any defense strategy, which largely remains invariant with changing nature of warfare (in this case size of data).

## iv) Big Data comes with Ethical Issues

- The ethical standards of the IC versus that of the public has always remained a historic tussle[18]. Recently the leaks by Edward Snowden on the NSA surveillance of large scale public data of the US citizens came under strong criticism and reinforced debates on privacy laws. Big data practices in IC cannot be viewed separately from the ethical standards that has been warranted in these debates. Eventually the IC will be compelled to uphold, what is being mentioned in literature as, *Big Data Ethics*: "—a social understanding of the times and contexts when big data analytics are appropriate, and of the times and contexts when they are not".[19]

- A survey on ethical use of big data conducted by the White House[20] (24,000 participants) revealed that majority had a strong lack of trust against the usage of big data particularly towards intelligence community. The report further noticed that most over 80% participants are concerned about transparency in the practices related to big data analysis.

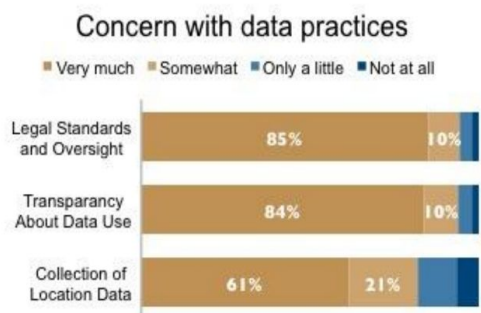

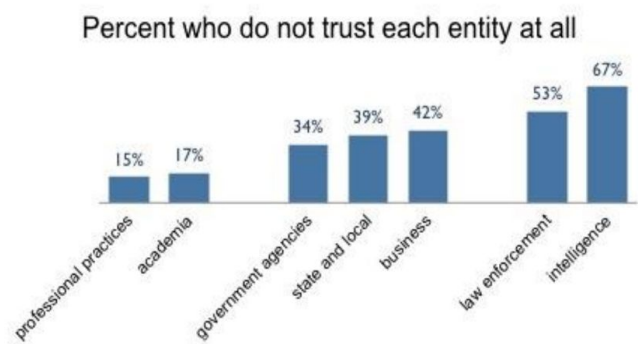

---

- Based on that, the three major ethical issues (or rather paradoxes) that limits role of big data sciences in IC are:[21]

    - **Transparency:** The big data being used by IC is mostly open source and hence transparent. But the analysis that is being done on them is opaque. Eventually, the public, who are the source of this open data, will warrant their right to know what decisions in IC are being affected by such open data, and may also question the ethical use of monitoring their life to form such judgements.

    - **Identity:** When faced with unverifiable source of information, IC will rely on big data analysis to uncover the identity of the source. But this very much threatens the identity, as the analysis can lead to absurd correlations of an identity with criminal intentions. This will end up forcing people to remain anonymous.

    - **Power:** The information from big data is prone to be misused for manipulation. Governments and dictatorial regimes will allow open source data to flourish so as to covertly restrict individuals that tend to be anti-establishment.

    - Examples:
    (i) During the Arab Spring, the Syrian government lifted the ban on social media sites, simply because it was easier to locate individuals and movements using the open source data and and analytics.

    (ii) The *Global Database of Events, Languages, and Tones*[22], has a record 250 million entries of worldwide conflicts since 1979. It performs a real time big data analysis on open source data and highlights regions of future conflicts. Such information can be used by authoritarian governments to tackle any possible conflicts.

The inherent issues of big data sciences thus affects the decision making process of intelligence community. It is quite important to reaffirm the role of intelligence community leaders, analysts and experts with domain knowledge for efficient usage of big data. Also, it is important to have a policy and framework in place so that future issues of big data can be sufficiently addressed.

In the next section I take a case study of nuclear intelligence to showcase the limitations of big data stated in this section, and provide a heuristic argument based on simple analysis

---

which leads to much specific questions from intelligence perspective, than the overdose of correlation from big data.

## 5.2 Case Study: Big Data & Nuclear Intelligence

*"We can't let our treaties get ahead of our monitoring and verification headlights."*

**--Defense Science Board Task Force Report**[23]

The role of big data has gained importance recently in the purview of nuclear intelligence. By analyzing real time data from multiple sensors, big data analysis can lead to sophisticated intelligence towards monitoring of nuclear deterrence, treaties and power plants. In 2014, a Task Force appointed by the Defense Science Board presented a report to the Office of The Secretary of Defense on "*Assessment of Nuclear Monitoring and Verification Technologies*".[24]

In one of the "paradigm shift" conclusions, the report stated:

- "Revamping the monitoring framework to identify proliferants early or well before the fact. The framework should: -
    - Adopt / adapt new tools for monitoring (e.g., **open and commercial sources, persistent surveillance from conventional war-fighting, "big data" analysis**) across the IC, DOD, and DOE"

The monitoring framework described in the report is a three stage process:

(i) **Access:** Mutual agreements for access from the of data from the nuclear sites. Any difficulty in access could be identified as potential proliferation.

(ii) **Sensing:** Having sensors to measure radiation from special nuclear material, satellite imagery, the "INTs". This will require technologies beyond agreed in current treaties.

(iii) **Assessment:** Analyzing the vast amount of data sets coming from stage (ii) using big data science and open source imagery and assess if any threat.

This cyclic monitoring framework, "Access-Sensing-Assess-Iterate" is mentioned in the figure below.

---

[23] *STOP LOOSE NUKES WITH BIG DATA AND CROWDSOURCING, EXPERTS URGE,* NextGov, 2014
http://www.nextgov.com/big-data/2014/01/stop-loose-nukes-big-data-and-crowdsourcing-experts-urge/77324/

[24]Defense Science Board Task Force Report. *Assessment of Nuclear Monitoring and Verification Technologies,* January 2014
http://www.acq.osd.mil/dsb/reports/NuclearMonitoringAndVerificationTechnologies.pdf

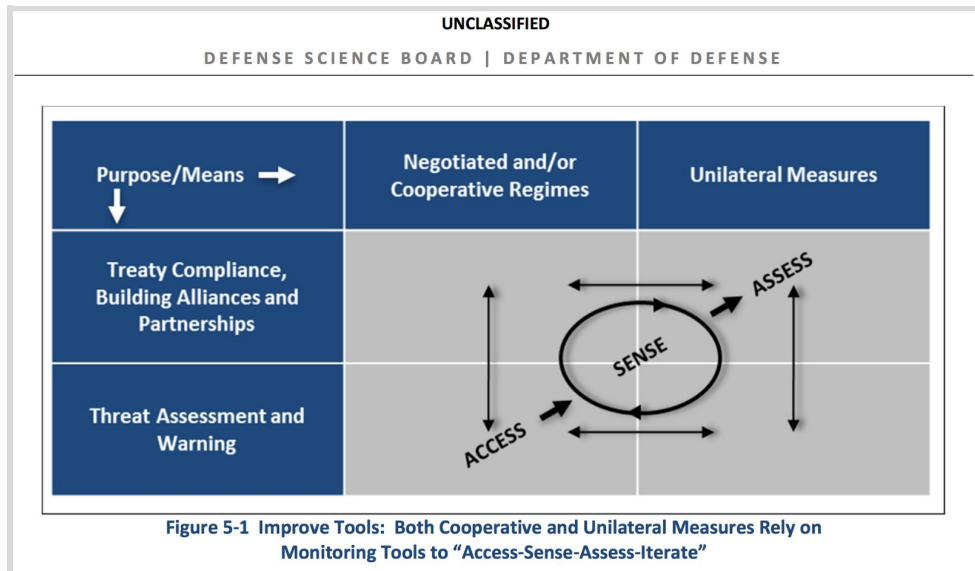

**Figure 5:** Nuclear Monitoring Cycle from Defense Science Board Task Force Report

The stage (iii) requires dedicated supercomputing infrastructure and softwares to analyze the information to be presented for quick follow up. The United States has strongly considered the investment in this area of supercomputing, and recently IBM and NVIDIA established a $425 Million facility at the Oak Ridge National Laboratory "to advance key research initiatives for national nuclear deterrence" [25].

However, in a January 2015 report by the World Institute for Nuclear Security, the experts group presented a warning for such investments[26]:

- *"Application of Big Data technology to nuclear security was probably premature and not warranted financially; the costs of significant Big Data solutions can run to $millions/annum and* **the perceived corporate benefits may not justify such investment at present***."*

- *"Should you rely on external analysts working for a contractor? – The general feeling was that this was not an ideal solution and that* **in-house expertise was desirable but would take time to develop***"*

---

In wake of such warnings, it is important to relook at the question on what else can we learn about nuclear intelligence from basic data analytics instead of getting carried away with buzzwords "Big Data and Supercomputers". In an attempt to understand the role of tacit knowledge in nations that nuclear powers, Adam Stulberg and his group[27] [28], examined the correlation between academic journals and the authors in the field of nuclear physics and engineering in Pakistan and Iran. This involved searching the academic database for set of keywords that are linked with nuclear weapons (plutonium, uranium etc). Each keyword is assigned a ranking and accordingly a score is assigned to the authors.

The correlations, then examined using visual analytics such as network diagrams, helped identify key individuals (as outliers) who played significant role in passing tacit knowledge. Along with that, the study also shows organizational pattern and decision making structure in those country. This provides insights in the inner workings of the government and back channels. Such studies are important for two reasons:

1. It utilizes open source data (academic journals), which is structured and is only few GBs in size.

2. The simple data analytics leads to very specific questions that can be directly addressed through experts with domain knowledge (why does Pakistan has nuclear intelligence structured around A. Q. Khan, while the name that comes for Iran is a rather lesser know individual?)

As a proof concept, if the same study is carried using all open source social media data, would it gain us further insight in nuclear intelligence? The answer is broadly no. But more significantly it can derail the strong conclusions that were already obtained through minimal computations. If one searches social media, say Twitter (petabytes of data), for the same keywords, it will at first order create a cross correlations with the worldwide trends, and those will likely be bizarre (currently "nuclear" search tag shows tweets about Trump). Clearly, the noise exceeds signal in Twitter type of database. This is a classic case on when not to use big data sciences.

As reported earlier by the Defense Science Board Task Force, the social media indicators can be used to monitor treaties and proliferations. If someone tweets from a region where there have been other independent INTs suggesting reports of violations then indeed such information is useful. But as the social media data is dynamic, the latency between analyzing relatively older data set and the new information coming in will make it difficult to mark any quick actions.

---

# 3. Conclusion and Future Work

The study reviews the scope of big data in context of intelligence community. There is no doubt that big data holds critical importance to a nation, in particular to preserve national security. The US government places central importance in the growth of technology, policies and ethical framework surrounding big data.

In scrutinizing the inherent issues in big data, I list four main limitations where it particularly affects the intelligence community:

(i) Reliance on experts for efficient use

(ii) Addressing knowledge gaps

(iii) Temporal use of current infrastructure

(iv) Ethical issues that will affect scope of analysis

In terms of hype around big data, we can conclude there is definite help in bridging information gaps but that doesn't make the domain knowledge irrelevant. In fact, for efficient use of big data in intelligence community, the issues stated above reaffirm the need for human input and experts to guidance.

As a case study on potential misuse of big data for intelligence community, I showcase the nuclear intelligence. I conclude that expensive infrastructure and over analyzing of heavy datasets would rather deviate us from conclusions that can be achieved by analytics on moderate sized open source academic data.

The study can be further strengthened by scrutinizing the effect of big data and its use in intelligence community towards:

a) Arab Spring and using social media as an indicator for predicting real world events[29]

b) Crowdsource forecasting tournaments to predict disruptive technologies[30]

c) Reliance of intelligence agencies on the private sectors for developing tools on big data sciences

---

[29] *Predicting The Future: Fantasy Or A Good Algorithm?.* NPR, 2012
http://www.npr.org/2012/10/08/162397787/predicting-the-future-fantasy-or-a-good-algorithm

[30] *IC embraces open source intel, even if it is double-edged.* Defense Systems, 2015
https://defensesystems.com/articles/2015/09/25/iarpa-open-source-intelligence.aspx?admgarea=DS

As an ending remark, I will like to point at the Hype cycle of emerging technologies in Figure-6. Until 2014, the big data / machine learning always remained at the peak inflated expectations, but with more studies such as this, there is fair understanding of the practical limitations yet reliance of big data sciences, and from 2015 it seems to be reaching more realistic expectations.

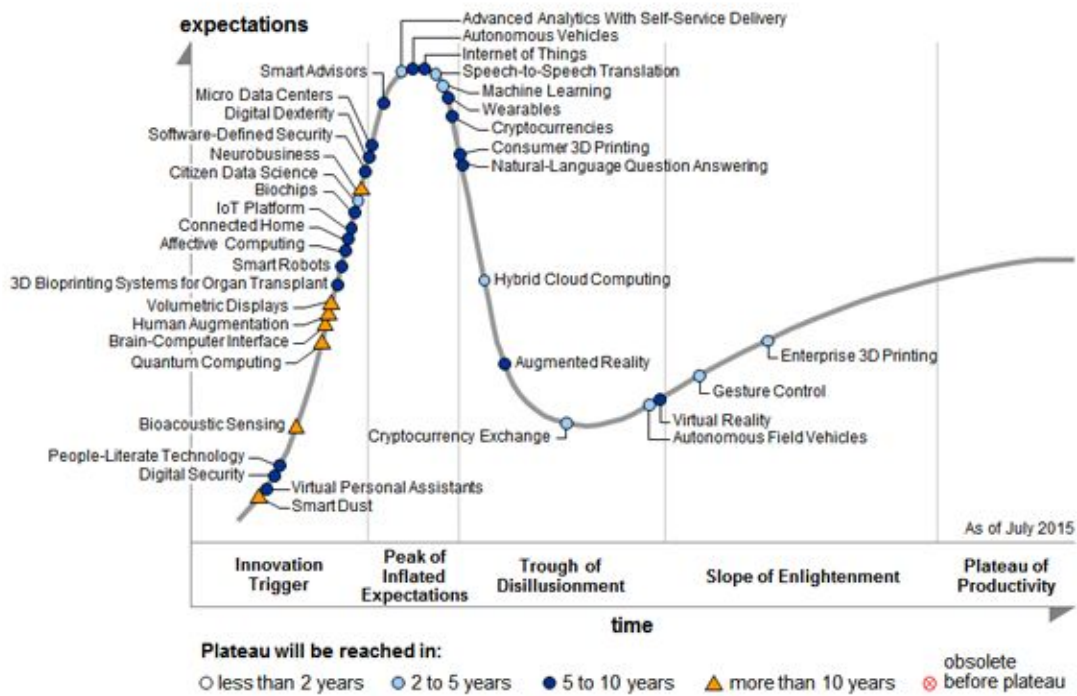

**Figure 6:** Hype Cycle of Emerging Technologies. The big data / Machine learning has barely crossed yet hype of expectations. [31]

# 5. Acknowledgements

The author will like to thank Margaret Kosal, Adam N. Stulberg and the *Sam Nunn Security Program* fellows for constructive feedback on this study.

---